# Comparative Analysis of Control Observer-based Methods for State Estimation of Lithium-ion Batteries in Practical Scenarios

Muhammad Saeed, *Student Member, IEEE*, Arash Khalatbarisoltani, *Member, IEEE*, Zhongwei Deng, *Member, IEEE*, Wenxue Liu, Faisal Altaf, *Member, IEEE*, Shuai Lu, *Member, IEEE*, and Xiaosong Hu, *Fellow, IEEE*.

*Abstract*—The reliability, lower computational complexity, and ease of implementation of control observers make them one of the most promising methods for the state estimation of Li-ion batteries (LIBs) in commercial applications. To pave their way, this study performs a comprehensive and systematic evaluation of four main categories of control observer-based methods in different practical scenarios considering estimation accuracy, computational time convergence speed, stability, and robustness against measurement uncertainties. Observers are designed using a $2^{nd}$ order equivalent circuit model whose observability against different scenarios is rigorously investigated to verify the feasibility of the proposed analysis. Established techniques then are validated against driving datasets and their comparative usefulness is evaluated using an experimental setup. The analysis also evaluates the adaptability of different techniques to EV field data. The results indicate better accuracy, stability, robustness, and faster convergence for the PI and PID, while the estimations of the Luenberger observers find it hard to converge against highly dynamic loadfiles. Moreover, this study also discusses the sensitivity of observer-based techniques to battery ohmic polarization and voltage-related measurement uncertainties. The most remarkable contribution of the proposed study lies in providing guidance for researchers when choosing the control observers for online state estimation of LIBs.

## I. INTRODUCTION

Lithium-ion batteries (LIBs) have emerged as the most realistic energy storage option for renewable energy to realize global emission targets set by the United Nations Framework Convention on Climate (UNFCCC) [1]. Especially the recent reductions in prices have helped them gain control of the electric vehicles (EVs) industry [2], [3]. However, due to their complex chemistry and potential safety-related issues, a smart battery management system (BMS) is required for safe and reliable operations [4], [5]. A BMS uses measurable cell parameters to estimate different internal states and knowledge of these states is then used to perform charge/discharge control, fault diagnosis, cell balancing, thermal management, and range prediction [6], [7], [8], [9], [10]. Among these states, the battery state of charge (SOC) is regarded as the most critical indicator. An inaccurate estimation of any cell's SOC can result in incorrect balancing of cells, possible deep-discharge or over-charge, reduction in operation range, capacity deterioration, or even permanent damage, and continuation of operations in such conditions might result in a thermal runaway or a possible fire incident [11]. Multiple algorithms are developed over time to estimate this critical battery state which are divided into the direct calculation approaches (DCAs), electrochemical model (EM)-based methods, equivalent circuit model (ECM)-based methods, and data-driven techniques (DDTs) [12].

The incompatibility of DCAs with real-time scenarios, the high computational complexity of EMs, and the lack of high-quality massive training data and generalizability of DDTs are the clear obstacles in the way to commercial applications [13], [14]. In this context, the ECM-based methods provide a perfect tradeoff between fidelity and complexity [12]. The ECMs use the resistor-capacitor (RC) networks to mimic the battery cell's electrochemical behavior and provide reasonable accuracy within a wide frequency range [15]. These methods are largely divided into filter-based and control observer-based techniques.

Filter-based methods mainly include the Kalman filters (KF) with their extensions [16], particle filters, and the combination of filters with learning algorithms [17]. The KF is a recursive process in which each step reallocates a trust weight of Ampere-hour-based and state-space model-based estimated SOC, and the final value is corrected by the Kalman gain [16]. The control observers, on the other hand, estimate the battery SOC based on the measurements of external parameters in a control system with state feedback. They provide high accuracy, robustness against model errors and uncertainties, enhanced control, and low complexity [5]. Both filters and observers-based techniques provide decent accuracy for SOC, and it's challenging to make a definitive recommendation. However, considering practical scenarios and the requirements of commercial applications, a case-specific comparison can be realized. In the context of large-scale BMS [18], [19], [20], using the published literature, the performance of both methods is reviewed against the model uncertainties, nonlinearities, reconfigurability, uncertainty of initial state, and computational complexity.

This work was supported in part by the National Key Research and Development Program of China (No. 2022YFB3305403), in part by the Talent Plan Project of Chongqing (No. cstc2021ycjh-bgzxm0295), and in part by the National Natural Science Foundation of China (52111530194). (Corresponding authors: Xiaosong Hu; Shuai Lu).

Muhammad Saeed and Shuai Lu are with the School of Electrical Engineering, Chongqing University, Chongqing 400044, China. (e-mail: msaeedch@cqu.edu.cn; shuai.lu@cqu.edu.cn)

Arash Khalatbarisoltani, Wenxue Liu, and Xiaosong Hu are with the College of Mechanical and Vehicle Engineering, Chongqing University, China. (e-mail: arash.khalatbarisoltani, wenxueliuu@cqu.edu.cn; xiaosonghu@ieee.org).

Zhongwei Deng is with the School of Mechanical and Electrical Engineering, UETC, Chengdu, China, 611731. (e-mail: dengzw1127@uestc.edu.cn).

Faisal Altaf is the battery Chief Engineer at the Electromobility Department, Volvo Group, Gothenburg, Sweden. (e-mail: faisal.altaf@volvo.com)

A comprehensive study [21] shows that the extended Kalman filter (EKF) gives a slightly better SOC accuracy with no model uncertainty; however, the estimations of EKF become unstable in the presence of model errors, while the nonlinear observer keeps performing well. Moreover, the study also reveals a high error in EKF estimations when the sensor noise covariance is considerably large. A recent study [22] shows that the errors of the designed observer are at least one order of magnitude smaller than the unscented Kalman filter (UKF) when unknown initial state conditions, ±30% parametric uncertainty, and model errors are considered. In addition, the vulnerability of UKF to high current dynamic profiles is also revealed. An observer given in [23] proves enhanced robustness against parameters and model errors when compared to EKF. Similarly, a study presented in [24] verifies the superiority of the designed nonlinear observer compared to EKF in practical scenarios. In addition, the observers offer much faster convergence against unknown initial SOC [22], [24], [25]. Furthermore, multiple works have verified a much lower computational complexity of control observers compared to KF-based techniques [25], [26].

In light of the above discussion, the observer-based methods have a clear edge over filters-based techniques for practical applications. Moreover, in BMS, each cell's SOC estimation is required to effectively balance all cells to realize enhanced range, prolonged life, and better safety for LIBs. Considering a very large-scale BMS, the lower computational complexity of control observers makes them a preferred choice. Furthermore, observers are the mature techniques [27], [28], [29], [30], and find great compatibility with already controller-packed battery energy storage system (BESSs) [31]. Till now, multiple studies have been reported to realize better accuracy, fast convergence, or enhanced robustness using a control observer [32], [33], [34], [35]. However, the comparative analysis, a vital tool to evaluate the most suitable method for a specific application, of different observers against practical scenarios of LIBs remained non-existent. Therefore, this analysis will help the researchers to realize the pros and cons of different observers when applied to battery state estimation and select the most suitable method according to their goal, nature of application, and design resources. Moreover, the comparative work done using real EV field data can provide valuable insights into the adaptability of different methods to a commercial application. A careful cell parameters sensitivity analysis can point out the weaknesses of observer-based battery SOC estimation techniques. In addition, the detailed analysis against measurement uncertainties can be quite helpful for professionals in identifying the most preferred method according to the accuracy of available battery sensors. To realize the proposed analysis systematically, this study first classifies them into the Luenberger observer [33], sliding mode observer [34], PI observer [32], and PID observer [5]. The main contributions can be summarized as follows:

1)- The established observer-based methods are designed using a 2nd order ECM system whose observability is rigorously discussed against different scenarios to verify the feasibility of the proposed analysis. 2)- The battery ECM is parameterized using improved experimental methods and the particle swarm optimization (PSO) technique and the designed observers are validated using multiple driving profiles. 3)- The comparative usefulness of different observers is assessed by an experimental setup considering different driving scenarios, critical operating temperatures, and unknown initial states. Moreover, real EV field data is used to test their compatibility with commercial applications. 4)- The resilience of different methods against cell ohmic resistance perturbations is also assessed after identifying ohmic resistance as the dominant parameter through a careful sensitivity analysis. 5)- Furthermore, multiple scenarios are set up to reveal the effectiveness of established techniques against sensor measurement errors, convergence speed in unexpected system shutdowns, and computational efficiency.

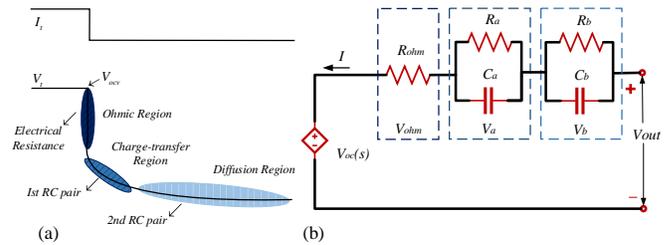

**Fig. 1.** (a) Impulse response of LIBs (b) Second order equivalent circuit model

The remainder of this paper is organized as follows. Section II details the selected battery ECM system and the design of the established observers. Section III shows experimental details, battery characteristic analysis battery cells, and validation of the designed observers against multiple driving datasets. Section IV discusses the results of the proposed comparative analysis. Section V summarizes the conclusions of this work.

## II. PROBLEM FORMULATION

### A. Battery Model

To better capture the complex dynamics of LIBs, this study considers the 2nd order ECM for the proposed research work. In the selected model, the ohmic polarization of LIB is represented by $R_{ohm}$. $R_a$ denotes the charge-transfer polarization of the cell, and $C_a$ represents the double-layer capacitance. Similarly, the concentration polarization and diffusion capacitances of the cell are represented by $R_b$ and $C_b$, respectively. In addition, the model includes a SOC-dependent open circuit voltage ($OCV$) source denoted by $V_{oc}(s)$, battery current $I$, and terminal voltage $V_{out}$, as illustrated in Fig. 1. By applying Kirchhoff's principle around the first RC pair of battery model, we can get:

$$I = I_{Ra} + I_{Ca} = \frac{V_a}{R_a} + C_a \frac{dV_a}{dt} \Rightarrow \dot{V}_a = \frac{-V_a}{R_a C_a} + \frac{I}{C_a} \quad (1)$$

Similarly, for the second RC pair, we can write:

$$I = I_{R_b} + I_{C_b} = \frac{V_b}{R_b} + C_b \frac{dV_b}{dt} \Rightarrow \dot{V}_b = \frac{-V_b}{R_b C_b} + \frac{I}{C_b} \quad (2)$$

As the instantaneous value of the SOC, denoted by $s(t)$, at a given time $t$ is the sum of the initial SOC "$s(0)$" and the amount of charge deposited or retrieved in that given time interval, the cell SOC can be depicted by the following expression:

$$s(t) = s(0) + \Delta s = s(0) + \int_0^t \frac{\eta I(\tau)}{C_t} d\tau \Rightarrow \dot{s} = \frac{\eta I}{C_t} \quad (3)$$

where $C_t$ is the total capacity and $\eta$ is coulombic efficiency of the cell. The terminal voltage $V_{out}$ can be written as:

$$V_{out} = V_{oc}(s) + V_{ohm} + V_a + V_b = V_{oc}(s) + IR_{ohm} + V_a + V_b \quad (4)$$

The battery OCV has a complicated nonlinear relationship with SOC. However, a linearized approximation ($Voc(s) = m_i s + c_i$, section III-27) can be realized experimentally. By combining it with (1)-(4), we get the state equation for the ECM system as:

$$\begin{cases} \dot{V}_a = -\frac{1}{R_a C_a} V_a + \frac{1}{C_a} I \\ \dot{V}_b = -\frac{1}{R_b C_b} V_b + \frac{1}{C_b} I \\ \dot{s} = \frac{\eta_i}{C_t} I \\ V_{out} = V_a + V_b + m_i s + c_i + IR_{ohm} \end{cases} \quad (5)$$

As the proposed work involves both linearized and non-linear practical scenarios, the linearized scenario is considered first by entirely ignoring the nonlinearities, disturbances, and measurement errors. By comparing (5) with the general state representation of a linearized system $\{\dot{x} = Ax + Bu, y = Cx + Du\}$, we get the following linearized state space representation:

$$\begin{cases} \begin{bmatrix} \dot{V}_a \\ \dot{V}_b \\ \dot{s} \end{bmatrix} = \begin{bmatrix} -1/R_a C_a & 0 & 0 \\ 0 & -1/R_b C_b & 0 \\ 0 & 0 & 0 \end{bmatrix} \begin{bmatrix} V_a \\ V_b \\ s \end{bmatrix} + \begin{bmatrix} 1/C_a \\ 1/C_b \\ \eta_i/C_t \end{bmatrix} I \\ V_{out} - c_i = \begin{bmatrix} 1 & 1 & m_i \end{bmatrix} \begin{bmatrix} V_a & V_b & s \end{bmatrix}^T + IR_{ohm} \end{cases} \quad (6)$$

where:

$$x = \begin{bmatrix} V_a \\ V_b \\ s \end{bmatrix}, A = \begin{bmatrix} -1/R_a C_a & 0 & 0 \\ 0 & -1/R_b C_b & 0 \\ 0 & 0 & 0 \end{bmatrix}, B = \begin{bmatrix} 1/C_a \\ 1/C_b \\ \eta_i/C_t \end{bmatrix}$$

$$u = I, C = \begin{bmatrix} 1 & 1 & m_i \end{bmatrix}, D = R_{ohm}, y = V_{out} - c_i$$

To evaluate the practical usefulness of different observers, the actual case of LIBs with nonlinearities, added disturbances, and measurement errors, can be depicted as follows:

$$\begin{cases} \dot{x} = Ax + Bu + Qf(u, x, t) \\ y = Cx + Du + w \end{cases} \quad (7)$$

where $w$ represents measurement errors and $f(x, u, t)$ describes the nonlinearities a function of control inputs, system states, or time. $Q$ matrix indicates the influence of these nonlinearities on different states of the system. By comparing (5) with (7), we get the following state space representation for the nonlinear case:

$$\begin{cases} \begin{bmatrix} \dot{V}_a \\ \dot{V}_b \\ \dot{s} \end{bmatrix} = \begin{bmatrix} -1/R_a C_a & 0 & 0 \\ 0 & -1/R_b C_b & 0 \\ 0 & 0 & 0 \end{bmatrix} \begin{bmatrix} V_a \\ V_b \\ s \end{bmatrix} + \begin{bmatrix} 1/C_a \\ 1/C_b \\ \eta_i/C_t \end{bmatrix} I + \begin{bmatrix} Q_1 \\ Q_2 \\ Q_3 \end{bmatrix} f \\ V_{out} - c_i - w = \begin{bmatrix} 1 & 1 & m_i \end{bmatrix} \begin{bmatrix} V_a & V_b & s \end{bmatrix}^T + IR_{ohm} \end{cases} \quad (8)$$

where:

$$x = \begin{bmatrix} V_a \\ V_b \\ s \end{bmatrix}, A = \begin{bmatrix} -1/R_a C_a & 0 & 0 \\ 0 & -1/R_b C_b & 0 \\ 0 & 0 & 0 \end{bmatrix}, B = \begin{bmatrix} 1/C_a \\ 1/C_b \\ \eta_i/C_t \end{bmatrix}, Q = \begin{bmatrix} Q_1 \\ Q_2 \\ Q_3 \end{bmatrix}$$

$$u = I, C = \begin{bmatrix} 1 & 1 & m_i \end{bmatrix}, D = R_{ohm}, y = V_{out} - c_i - w$$

*B. Observability of LIBs through selected ECM system*

The demonstration of complete observability for the selected ECM is essential to verify that the internal states of the LIB system in each scenario can be estimated using a state observer.

*1) Linearized Scenario*

As the linearized battery ECM model (6) is not augmented, controllability exits. So, the system is considered completely observable for every $x(t_0)$ if both the controllability matrix and observability matrix are full rank [37]. The controllability and observability matrices for the linearized scenario (6) of ECM can be written as follows:

$$Con = \begin{bmatrix} B \\ AB \\ A^2 B \end{bmatrix} = \begin{bmatrix} 1/C_a & -1/R_a C_a^2 & 1/R_a^2 C_a^3 \\ 1/C_b & -1/R_b C_b^2 & 1/R_b^2 C_b^2 \\ \eta_i/C_t & 0 & 0 \end{bmatrix} \quad (9)$$

$$O = \begin{bmatrix} C \\ CA \\ CA^2 \end{bmatrix} = \begin{bmatrix} 1 & 1 & m_i \\ -1/R_a C_a & -1/R_b C_b & 0 \\ 1/R_a^2 C_a^2 & 1/R_a^2 C_a^2 & 0 \end{bmatrix} \quad (10)$$

As the values of terms $\eta/C_t$, $1/C_a$, $1/C_b$, $1/R_a C_a$, $-1/R_b C_b$, and $m_i$ cannot be zero during any practical condition of LIBs, both matrices are full ranks, and the linearized system will remain observable for every $x(t_0)$.

*2) Non-linear Scenarios*

The rigorous observability criterion for the nonlinear practical scenarios is usually settled down using the tools of the Lie derivatives. According to the Lie derivatives observability theorem [38], the system of the following form:

$$\begin{cases} \dot{x} = f(x) + gu \\ y = h(x) + Ku \end{cases} \quad (11)$$

is observable at any $x_0$, if the following matrix of the gradients of the Lie derivative is full rank.

$$O = \begin{bmatrix} dh & dL_f^1 h & dL_g^1 h & dL_f^2 h & dL_g^2 h & . & . \end{bmatrix}^T \quad (12)$$

where:

$$dh = \partial h / \partial x = \begin{bmatrix} \partial h / \partial x_1 & \partial h / \partial x_2 & . & . & \partial h / \partial x_n \end{bmatrix}$$

$$L_f h(x) = \partial h / \partial x . f(x) = \begin{bmatrix} \partial h / \partial x_1 & \partial h / \partial x_2 & . & . & \partial h / \partial x_n \end{bmatrix} f(x)$$

$$L_g h(x) = \partial h / \partial x . g(x) = \begin{bmatrix} \partial h / \partial x_1 & \partial h / \partial x_2 & . & . & \partial h / \partial x_n \end{bmatrix} g(x)$$

$$L_f^k h(x) = L_f (L_f^{k-1} h(x))$$

$$L_g^k h(x) = L_g (L_g^{k-1} h(x))$$

First, by investigating the scenario of input nonlinearities in LIBs ECM system, and re-writing the prescribed nonlinear scenario into form (11), we can the following representations:

$$\begin{cases} \begin{bmatrix} \dot{V}_a \\ \dot{V}_b \\ \dot{s} \end{bmatrix} = \begin{bmatrix} -1/R_a C_a & 0 & 0 \\ 0 & -1/R_b C_b & 0 \\ 0 & 0 & 0 \end{bmatrix} \begin{bmatrix} V_a \\ V_b \\ s \end{bmatrix} + \begin{bmatrix} 1/C_a + Q_1 \\ 1/C_b + Q_2 \\ \eta_i/C_t + Q_3 \end{bmatrix} u \\ V_{out} = V_{oc}(s) + V_a + V_b + uR_{ohm} \end{cases} \quad (13)$$

where, $x=[V_a, V_b, s]^T$, $u=I$, $y=V_{out}$, $f(x)=[(-1/R_a C_a).V_a\ (-1/R_b C_b).V_b\ 0]^T$, $g(x)=[(1/C_a+Q_1)\ (1/C_b+Q_2)\ (\eta_i/C_t + Q_3)]^T$, $h(x)=V_a+ V_b + V_{oc}(s)$ and $K=R_{ohm}$. Computing the gradients of the Lie derivative for (13), we can get the following matrix:

$$\begin{bmatrix} dh \\ dL_f^1h \\ dL_f^2h \\ dL_f^kh \\ dL_g^{k-1}h \end{bmatrix} = \begin{bmatrix} 1 & 1 & \frac{dVoc(s)}{ds} \\ \frac{1}{-R_aC_a} & \frac{1}{-R_bC_b} & 0 \\ \frac{1}{(-R_aC_a)^2} & \frac{1}{(-R_bC_b)^2} & 0 \\ \frac{1}{(-R_aC_a)^k} & \frac{1}{(-R_bC_b)^k} & 0 \\ 0 & 0 & (\frac{\eta_i}{C_t}+Q_3)^{k-1}\frac{d^kVoc(s)}{ds^k} \end{bmatrix} \quad (14)$$

Now considering the nonlinearities as a function of the states, and transforming the non-linear scenario of LIBs into (11), we get: $x=[V_a, V_b, s]^T$, $u=I$, $y=V_{out}$, $f(x)=[(-1/R_aC_a+Q_1).V_a$ $(-1/R_bC_b+Q_2).V_b$ $Q_3]^T$, $g(x)=[1/C_a$ $1/C_b$ $\eta_i/C_t]^T$, $h(x)=V_a+V_b+V_{oc}(s)$ and $K=R_{ohm}$. Again, by computing the gradients of the Lie derivative for it, we get the following matrix:

$$\begin{bmatrix} dh \\ dL_f^1h \\ dL_f^2h \\ dL_f^{k-1}h \\ dL_g^{k-1}h \end{bmatrix} = \begin{bmatrix} 1 & 1 & \frac{dVoc(s)}{ds} \\ \frac{-1}{R_aC_a}+Q_1 & \frac{-1}{R_bC_b}+Q_2 & Q_3\frac{d^2Voc(s)}{ds^2} \\ (\frac{-1}{R_aC_a}+Q_1)^2 & (\frac{-1}{R_bC_b}+Q_2)^2 & Q_3^2\frac{d^3Voc(s)}{ds^3} \\ (\frac{-1}{R_aC_a}+Q_1)^{k-1} & (\frac{-1}{R_bC_b}+Q_2)^{k-1} & Q_3^{k-1}\frac{d^kVoc(s)}{ds^k} \\ 0 & 0 & (\frac{\eta_i}{C_t})^{k-1}\frac{d^kVoc(s)}{ds^k} \end{bmatrix} \quad (15)$$

Taking the nonlinearities in terms of measurement errors $w$, which will result in $y=V_{out}-w$. Then transforming this scenario into (11), we get: $x=[V_a, V_b, s]^T$, $u=I$, $y=V_{out}-w$, $f(x)=f(x)=[(-1/R_aC_a).V_a$ $(-1/R_bC_b).V_b$ $0]^T$, $g(x)=[1/C_a$ $1/C_b$ $\eta_i/C_t]^T$, $h(x)=V_a+V_b+V_{oc}(s)+$ and $K=R_{ohm}$. Again, by using the above expression, and computing the gradient of the Lie derivative for it, we get:

$$\begin{bmatrix} dh \\ dL_f^1h \\ dL_f^2h \\ dL_f^kh \\ dL_g^{k-1}h \end{bmatrix} = \begin{bmatrix} 1 & 1 & \frac{dVoc(s)}{ds} \\ \frac{1}{-R_aC_a} & \frac{1}{-R_bC_b} & 0 \\ \frac{1}{(-R_aC_a)^2} & \frac{1}{(-R_bC_b)^2} & 0 \\ \frac{1}{(-R_aC_a)^k} & \frac{1}{(-R_bC_b)^k} & 0 \\ 0 & 0 & (\frac{\eta_i}{C_t})^{k-1}\frac{d^kVoc(s)}{ds^k} \end{bmatrix} \quad (16)$$

It can be seen that the observability matrices of (14), (15), and (16) are full rank if the term $d^kV_{oc}(s)/ds^k \neq 0$. As the battery OCV is a nonlinear function of the SOC, the $d^kV_{oc}(s)/ds^k \neq 0$ will stand for every operating situation, and all matrices will be full rank. As the studied battery ECM system is proved observable in every operating scenario, the system's internal states can be estimated using a nonlinear observer. Hence, a comparative analysis of different control observer-based state estimation methods in practical scenarios of LIBs is feasible.

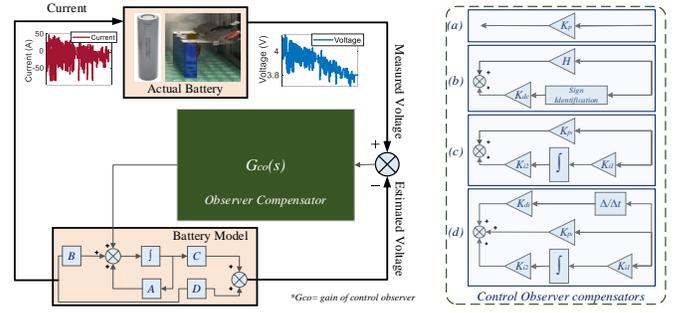

**Fig. 2.** General representation of observers-based SOC estimation methods; Observer compensator (a) Luenberger, (b) Sliding mode, (c) PI, and (d) PID

### C. Control Observers-based Estimation Methods

The observer-based methods work by comparing the model-based estimated cell voltage with the measured voltage of the real battery cell. The difference of both values (error) is fed into the observer compensator block which after analyzing different aspects of the error generates a response signal. This error compensating response is added back into the system, as shown in Fig. 2. In the next cycle, the new voltage value is estimated based on the model plus the observer's compensating response. The system error is sent again to the observer compensator, which modifies the response depending on the new error. The modified response is added back into the system, and this whole cycle continues. The unknown or the required to-be-estimated parameters are selected as the system states while designing the observer. According to the nature of their control mechanism, the observer-based SOC estimations method can be divided into the following four categories.

*1. Luenberger Observer*

The Luenberger observer is based on the *P*-control only, and a wide range of processes and applications use this observer due to the simplicity of its working principle. It was introduced for state estimation of lithium-ion batteries in [33]. The working of the Luenberger observer can be expressed using the following mathematical expression where *L* shows the Luenberger gain.

$$\begin{cases} \dot{\tilde{x}} = A\tilde{x} + Bu + L(y-\tilde{y}) \\ \tilde{y} = Cx + Du \end{cases} \quad (17)$$

*2. Sliding Mode Observer*

Kim first brought the idea of using sliding mode (SM) observer into the field of state estimation for LIBs [39]. This observer employs a dis-continuous feedback sgn ($e_y$) in addition to the Luenberger observer's p-control. The mathematical formulation of the SM observer principle is as follows:

$$\begin{cases} \dot{\tilde{x}} = A\tilde{x} + Bu + H(y-\tilde{y}) + K_{dc}.\text{sgn}(y-\tilde{y}) \\ \tilde{y} = Cx + Du \end{cases} \quad (18)$$

where $H$ and $K_{dc}$ represent the gain matrix and switching gain, respectively. The dis-continuous feedback of the SM observer can be expressed as follows:

$$\text{sgn}(e_y) = \begin{cases} +1, e_y > 1 \\ -1, e_y < 1 \end{cases} \quad (19)$$

*3. Proportional Integral Observer*

The PI observer was introduced recently in the field of LIBs for state estimation and other control-related purposes [18],

[40]. It is fundamentally based on the Luenberger and includes the integral feedback part for better error convergence. The working of a PI observer can be depicted as follows:

$$\begin{cases} \dot{\tilde{x}} = A\tilde{x} + Bu + K_p(y - \tilde{y}) + K_{i2}C_{PI} \\ \dot{C}_{PI} = K_{i1}(y - \tilde{y}) \\ \tilde{y} = Cx + Du \end{cases} \quad (20)$$

where $K_p$ and $K_{i1}$ are the proportional and integral gain matrices, respectively. $C_{PI}$ is $K_{i2}$ times integration of the error signal.

### 4. Proportional Integral Derivative Observer

The derivative component of the PID observer helps achieve faster convergence during state estimation of LIBs [5]. However, this improvement normally comes at a cost. As the derivative feature is used to tackle the issues related to the rate of change of error in the system, the presence of very rapidly changing disturbances can initiate an aggressive derivative response, which can deteriorate the performance of the system. This control observer includes additional derivative and integral feedback parts for better error convergence, and an enhanced degree of freedom for applying adaptive control laws. The working of the PID observer is portrayed as follows:

$$\begin{cases} \dot{\tilde{x}} = A\tilde{x} + Bu + K_p(y - \tilde{y}) + K_{i2}C_{PI} + K_d E_d \\ \dot{C}_{PI} = K_{i1}(y - \tilde{y}) \\ E_d = \dfrac{d}{dt}(y - \tilde{y}) \\ \tilde{y} = Cx + Du \end{cases} \quad (21)$$

Where the $K_p$, $K_{i1}$, and $K_d$ are the proportional, integral, and derivative gain matrices, respectively. $E_d$ denotes the derivative of the system error. $C_{PI}$ is $K_{i2}$ times integration of error signal.

### D. Convergence and Observer Design

The convergence requirements for each observer's design can be studies through its state error response of the system. For it, considering the Luenberger and sliding mode observer first and denoting $e = \tilde{x} - x$, the following error system can be derived from (17) and (8), and (19) and (8), respectively.

$$\dot{e} = A_e e - Qf \quad (22)$$

where, the error matrix $A_e$ for Luenberger observer is:

$$A_e = \begin{bmatrix} -1/R_aC_a - L_1 & -L_1 & -L_1 m_i \\ -L_2 & -1/R_bC_b - L_2 & -L_2 m_i \\ -L_3 & -L_3 & -L_3 m_i \end{bmatrix}$$

Similarly, the error matrix $A_e$ for the sliding mode observer can be written as follows:

$$A_e = \begin{bmatrix} -1/R_aC_a - H_1 & -H_1 & -H_1 m_i \\ -H_2 & -1/R_bC_b - H_2 & -H_2 m_i \\ -H_3 & -H_3 & -H_3 m_i \end{bmatrix}$$

Similarly, by denoting $e = \tilde{x} - x$ and $\mathcal{R} = [e \; C_{PI}]^T$, the following error system for the PI and PID observer can be derived from the sets of (8) & (20), and (8) & (21).

$$\dot{\mathcal{R}} = A_e \mathcal{R} - Qf \quad (23)$$

where the error matrix $A_e$ for the PI observer can be written as:

$$A_e = \begin{bmatrix} -1/R_aC_a - K_{p1} & -K_{p1} & -K_{p1} m_i & K_{i2_1} \\ -K_{p2} & -1/R_bC_b - K_{p2} & -K_{p2} m_i & K_{i2_2} \\ -K_{p3} & -K_{p3} & -K_{p3} m_i & K_{i2_3} \\ -K_{i1} & -K_{i1} & -K_{i1} m_i & 0 \end{bmatrix}$$

Similarly, the error matrix $A_e$ for the PID observer is:

$$A_e = \begin{bmatrix} \dfrac{-1}{R_aC_a} - K_{p1} - \dfrac{K_{d1}}{\Delta t} & -K_{p1} - \dfrac{K_{d1}}{\Delta t} & (-K_{p1} - \dfrac{K_{d1}}{\Delta t})m_i & K_{i2_1} \\ -K_{p2} - \dfrac{K_{d2}}{\Delta t} & \dfrac{-1}{R_bC_b} - K_{p2} - \dfrac{K_{d2}}{\Delta t} & (-K_{p2} - \dfrac{K_{d2}}{\Delta t})m_i & K_{i2_2} \\ -K_{p3} - \dfrac{K_{d3}}{\Delta t} & -K_{p3} - \dfrac{K_{d3}}{\Delta t} & (-K_{p3} - \dfrac{K_{d3}}{\Delta t})m_i & K_{i2_3} \\ -K_{i1} & -K_{i1} & -K_{i1} m_i & 0 \end{bmatrix}$$

For each observer, the matric $A_e$ could be arbitrarily assigned if and only if the system is observable. Since the observability is proved in Section II-B, the gain parameters of each observer can be selected using the LQ method or the pole place method to assure $A_e$ is Hurwitz, indicating that the system would converge. Hence, we can conclude that $e \rightarrow 0$ and $C_{PI} \rightarrow 0$ as t $\rightarrow \infty$, meaning the estimated states of all observers would converge to the true states of the systems. The general working of the observer-based SOC estimation method is brief in Algorithm I.

## III. EXPERIMENTAL ARRANGEMENTS AND BATTERY CHARACTERISTIC ANALYSIS

### A. Experimental Arrangements

The large-format prismatic LIB cell *CALB-L148N58A* and Tesla cylindrical cell *INR21700-M50T* are selected for the experimentational work. The key parameters of both cells are listed in Table I. The experimental setup includes a Neware battery performance tester *CT-8016-5V100A-NTFA* to test the battery cells as per actual driving loadfiles, a temperature and humidity control chamber *DHT-375-40-AR-SD* to maintain the desired operating conditions, a data acquisition unit *CA-4008-IU-VT-TX*, and a host PC. The established experimental setup is illustrated in Fig. 3, while the technical specifications of the

---

**Algorithm I**: Control observers-based state of charge estimation of LIBs
**State Space Representation:**
Ref. to Eq. (8)
**Inputs:** Battery current $I$, measured voltage $V_{out}$,
battery parameters $R_{ohm}$, $R_a$, $C_a$, $R_b$, $C_b$
**Output**: Estimated SOC $\hat{x}_3$
**Initialization**: $K_{dc}$ $\hat{x}_j$ $L_j$, $H_j$, $K_{pj}$, $K_{ij}$, $K_{dj}$ for $j=1,2,3$
**Computations:**
- Compute the state vector $\hat{x}_k$ using:
(a) Luenberger Observer: using Eq. (17)
(b) Sliding mode observer: using Eq. (18)
(c) PI Observer: using Eq. (20)
(d) PID Observer: using Eq. (21)
Compute tracking target: $\hat{y}_k = C\hat{x}_k + Du_k \rightarrow \hat{V}_{out,k} = \hat{y}_k + c_k$
Update error: $e_k = V_{out,k} - \hat{V}_{out,k}$
Update error integral: $C_{PI,k} = C_{PI,k-1} - e_k \Delta t$
- Based on the above calculation and nature of error, calculate the error compensation response of each observer as follows:
Luenberger Observer: $[L_1 \; L_2 \; L_3]^T e_k$
Sliding mode Observer: $[H_1 \; H_2 \; H_3]^T e_k + K_{dc}. sgn(e_k)$
PI Observer: $[K_{p1} \; K_{p2} \; K_{p3}]^T e_k + [K_{i1} \; K_{i2} \; K_{i3}]^T C_{PI,k}$
PID Observer:
$[K_{p1} \; K_{p2} \; K_{p3}]^T e_k + [K_{i1} \; K_{i2} \; K_{i3}]^T C_{PI,k} + [K_{d1} \; K_{d2} \; K_{d3}]^T \dfrac{e_k}{\Delta t}$
**Return** Estimated SOC $\hat{x}_3$
**Step 5.** for next data value, return to the start of computation process

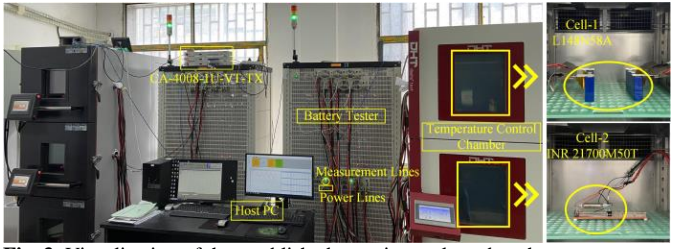

**Fig. 3.** Visualization of the established experimental test bench

TABLE I
KEY PARAMETERS OF SELECTED BATTERY CELLS

| Specification | CALB-L148N58A | INR-21700M50T |
|---|---|---|
| Total Capacity | 58 Ah | 4.8 Ah |
| Cathode | NMC 811 | $LiNiMnCoO_2$ |
| Nominal Voltage | 3.7 V | 3.63 V |
| Voltage range | 2.75-4.35 V | 2.5-4.2 V |
| Temperature range | -20~55ºC | -20~55 ºC |
| Weight | 926 g | 70 g |
| Dimension(mm) | 148.24x26.6x105.9 | D 21.44 x 70.80 |

utilized equipment are listed in Table II. In accordance with the *GB/T-31467* standards, screening and capacity calibration tests are performed before any experimentations to only select the cells whose capacity fluctuations are less than 3%.

### B. Characteristic analysis of LIB Cells

*1. OCV-SOC Relationship*

In this work, the low current test is used for identifying the battery OCV-SOC relationship. Both *L148N58A* and *INR21700-M50T* cells are first fully charged using the constant current constant voltage (CCCV) protocols. These fully charged cells are then discharged using a low current of C/30 till the specified $V_{min}$ (2.75V for prismatic cell and 2.5V for the cylindrical cell) arrives. After being rested for 2 hours, the battery cells are again charged using the same low current profile till their upper voltage limit. As a very low dis/charge weakens the internal polarizations and hysteresis effect to the greatest extent, the cell voltage during charging $V_{out(c)}$ can be calculated as follows:

$$V_{out(c)} = V_{oc}(s) + V_{ohm} + I(R_a + R_b) \quad (24)$$

Similarly, the cell voltage during discharging $V_{out(d)}$ equals to:

$$V_{out(d)} = V_{oc}(s) - V_{ohm} - I(R_a + R_b) \quad (25)$$

Now, the OCV can be achieved by averaging both curves.

$$V_{oc}(s) = (V_{out(d)} + V_{out(c)})/2 \quad (26)$$

The used low current test profiles and the OCV calculation mechanism are illustrated in Fig. 4. After achieving an accurate value of OCV at every data point, the following relationship can be formulated inside every two consecutive data points:

TABLE II
KEY SPECIFICATION OF EXPERIMENTAL EQUIPMENT

| Item Name | Type | Specification |
|---|---|---|
| Battery Performance tester | CT-8016-5V100A-NTFA | Range: 0.025V~5V; 0.5~100A (± 0.05% of accuracy on full scale) Resolution: AD/DA: 16 bits |
| Humidity & temperature control chamber | DHT-375-40-AR-SD | Cooling rate: 2 ºC/min Heating rate: 4 ºC/min Range: -40~150ºC (fluctuations<±0.5) |

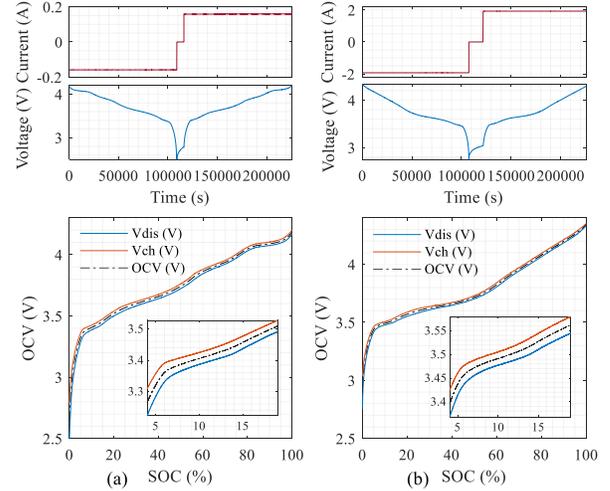

**Fig. 4.** OCV-SOC identification (a) INR21700M50T, and (b) L148N58A cell

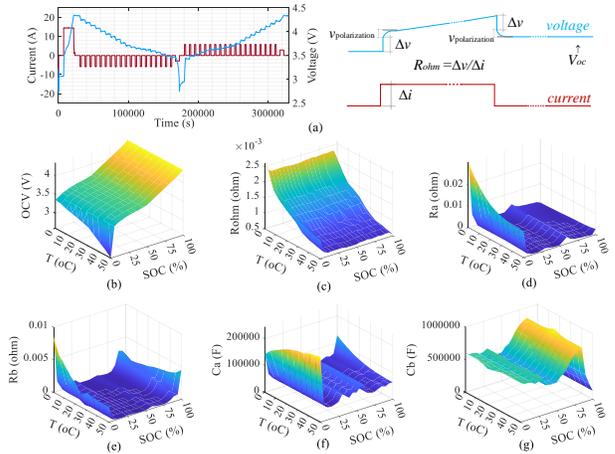

**Fig. 5.** (a) Explanation of the pulse test; map of cell parameters for *L148N58A* against the temperature and SOC: (b) OCV (c) $R_{ohm}$ (d) $R_a$ (e) $R_b$ (f) $C_a$ (g) $C_b$

$$V_{oc}(s) = m_i s + c_i \quad (27)$$

where: $m_i = \dfrac{V_{oc_{i+1}} - V_{oc_i}}{s_{i+1} - s_i}$ , and $c_i = \dfrac{V_{oc_i} s_{i+1} - V_{oc_{i+1}} s_i}{s_{i+1} - s_i}$

*2. Battery Parameters Identification*

For accurate identification of battery parameters, a specific hybrid power characterization (HPPC) test is performed at multiple temperatures, and a sample of the pulse test at 10ºC is illustrated in Fig. 5. The resulting battery data is processed by the particle swarm optimization (PSO) technique to compute the value of the cell parameters. The maps of identified battery parameters for *L148N58A* cells are illustrated in Fig. 5.

### C. Validation of control-observer against driving datasets

To validate the estimation capability of established observer-based methods, the experimentally collected driving datasets of the supplemental federal test procedure EPA US06 (SFTP-US06) and Beijing dynamic stress test (BJDST) profiles are provided as input to each observer. The estimations of each observer are compared with the measured values, and the results are shown in Fig. 6. It is clear from the results that all observer-based state estimation methods have the potential to track the measured cell voltage which verifies their ability for the state estimation of LIBs during practical scenarios.

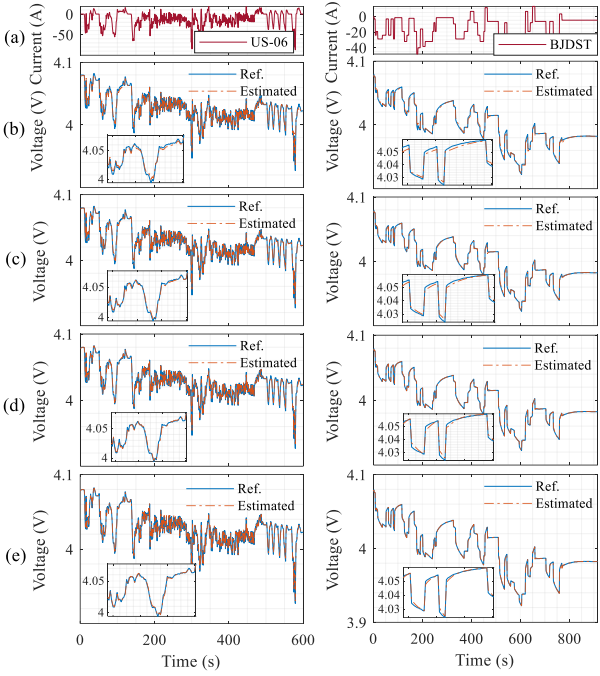

**Fig. 6.** (a) loadfiles; Validation results (b) Luenberger (c) SMO (d) PI (e) PID

## IV. RESULTS AND DISCUSSIONS

The comparative usefulness of all the observers is evaluated using the experimental setup (see Fig. 3). To further analyze their practical effectiveness in comparison with filter-based techniques, the square root cubature Kalman filter (SRCKF) is designed and tested alongside the established observers. All the techniques are designed on the host PC using the SIMULINK environment. The battery cells are placed in the temperature chamber in a desired environment. Using the battery tester, LIB cells are subjected to dynamic current loadfiles. The measured data of the cell current, voltage, and temperature is transferred to the host PC using TCP/IP commutations. The LIB's internal states are estimated using the received data, and results are compared with the measured values of these states to statistically evaluate the performance of each estimator using the following indices of root mean square error (*RMSE*), mean absolute error (*MAE*), and maximum absolute error (*MaxAE*).

$$\begin{cases} MaxAE = \max(|x_{o_i} - \hat{x}_{o_i}|) \\ RMSE = \sqrt{\frac{1}{N} \sum_{i=1}^{N} (x_{o_i} - \hat{x}_{o_i})^2} \\ MAE = \frac{1}{N} \sum_{i=1}^{N} |x_{o_i} - \hat{x}_{o_i}| \end{cases} \quad (28)$$

### A. Estimation Accuracy

To compare the estimation accuracy of different techniques against dynamic conditions, the LIB cells are subjected to the BJDST current profile starting at 80% of SOC, and the cell data is provided to the host PC using TCP/IP protocol. To focus only on the accuracy, all the estimators are provided with the correct initial state. The SOC estimation results of each method, with the prescribes error indices, are shown in Fig. 7. It is clear from the results that every technique shows a strong capability of tracking the reference SOC; however, the results exhibit a great manifestation of different control features. The Luenberger and SM observer show higher errors, and the absence of an integral feature refrains them from a perfect convergence. Similarly, the derivative feature of the PID observer results ensure better convergence against high nonlinearities as compared to the PI observer. Moreover, due to the known initial state value, the SRCKF performed better than Luenberger and SMO, while PID observer still outperformed SRCKF's estimations.

To intensify the analysis, a composite current sequence made of the federal urban driving schedule (FUDS), dynamic stress test (DST), STFP-US06, and BJDST is compiled using the battery tester. The cells are placed under a critical temperature of 45°C, and subjected to the composite current sequence. The cell data is provided to the designed observers using TCP/IP communications to predict the battery's internal state. A similar experiment is repeated at the low temperature of 5°C, and the estimation results of both experiments are plotted in Fig. 8.

It can be seen from the results that all estimators are capable of tracking the real SOC even at critical temperatures; however,

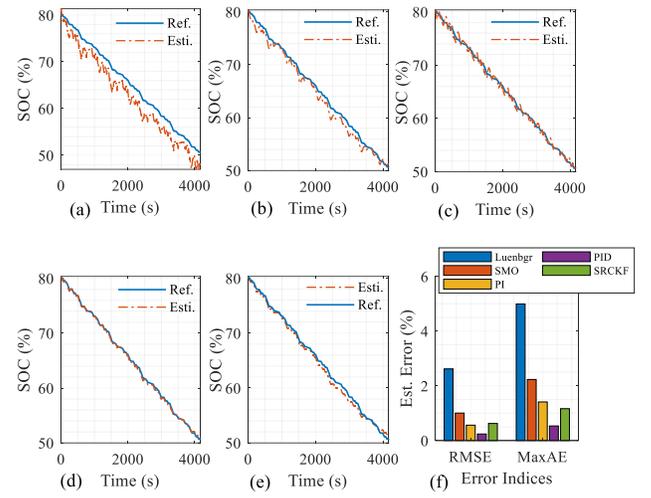

**Fig. 7.** SOC estimation results (a) Luenberger observer (b) SMO (c) PI observer (d) PID observer (e) SRCKF, and (f) statistical error indices of all techniques

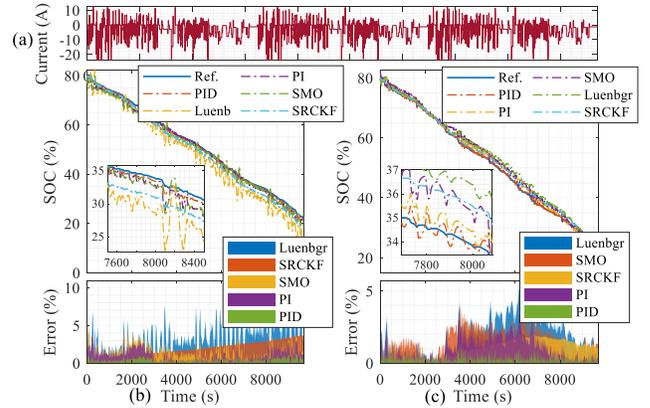

**Fig. 8.** (a) Used current sequence (b) SOC estimation results (c) estimation error

TABLE III
STATISTICAL ERROR INDICES FOR DIFFERENT TECHNIQUES

| Error | Luenberger | | SMO | | PI | | PID | | SRCKF | |
|---|---|---|---|---|---|---|---|---|---|---|
| (%) | 5°C | 45°C | 5°C | 45°C | 5°C | 45°C | 5°C | 45°C | 5°C | 45°C |
| *MaxAE* | 10.8 | 5.67 | 5.1 | 3.9 | 3.7 | 2.8 | 1.5 | 1.35 | 5.2 | 2.59 |
| *RMSE* | 4.8 | 3.31 | 2.3 | 1.98 | 1.9 | 1.22 | 0.8 | 0.68 | 2.2 | 1.76 |

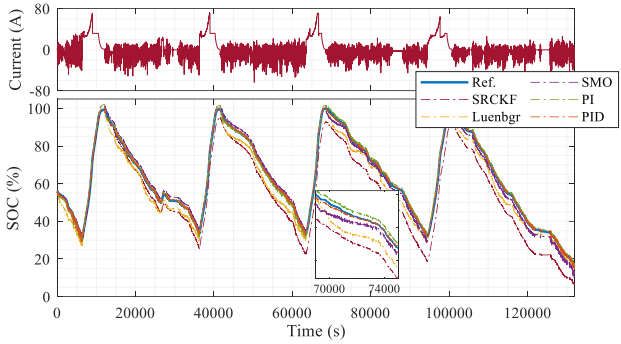

**Fig. 9.** EV field data (a) Current sequence, and (b) SOC estimation results

TABLE IV
STATISTICAL ERROR INDICES FOR DIFFERENT TECHNIQUES

| Error Indices | SRCKF | Luenberger | SMO | PI | PID |
|---|---|---|---|---|---|
| MaxAE | 13.44 | 8.9741 | 6.9872 | 4.8201 | 3.7944 |
| RMSE | 7.43 | 5.3076 | 2.9420 | 1.9743 | 1.2104 |

the accuracy of some observers deteriorated considerably. It is obvious from Table III that the Luenberger exhibits the highest deterioration, while the PID offers the highest resilience against critical temperatures. Due to the presence of modeling errors at critical conditions, the estimations of SRCKF started to drift away and become worse than the PI observer which verifies the better robustness of control observers in practical scenarios. Moreover, it is worth noticing that the high temperatures have minimal effect on observers' accuracy.

Finally, a winter segment of real EV data from the state key laboratory of intelligent vehicle safety technology (Chongqing) is used to evaluate the compatibility of all techniques with commercial applications. Validation against real EV field data is important because most algorithms diverge against on-road driving patterns, imperfect sensing, and aggressive dis/charging behaviors. A segment of 4 dis/charge cycles sampled every 10s is used, and the estimation results of each technique are shown in Fig. 9 and Table IV. It can be seen that all observers have a strong potential for SOC estimation in real-time with sufficient accuracy. Among them, the Luenberger observer suffers higher errors but it still can be used in low-end applications due to its simplicity. Most importantly, all observers converge to real state; however, the estimations of SRCKF tend to diverge with time because of modeling deficiencies of ECMs against real-time scenarios and unrealistic assumptions of Gaussian noise.

### B. Convergence Analysis

The fast convergence is a highly desirable feature in practical BMS applications, especially in case of unexpected shutdowns without saving prior data. To evaluate the prescribed scenario, all techniques are designed with no prior knowledge of initial SOC and are tested when LIB has different values of the initial SOC. In this study, 100% and 60% are selected as two initial values, and the time taken by each estimator to converge to the reference value is monitored. From the comparative results of Fig. 10, it is evident that all techniques can recover from initial high SOC error, and their convergence time differs a lot. The PID observer leads the others because it is equipped with both the derivative and integral features, which provide the fastest response against such scenarios. The PI observer and SRCKF fall behind closely, while the Luenberger and SM observers show a much slower response against such scenarios.

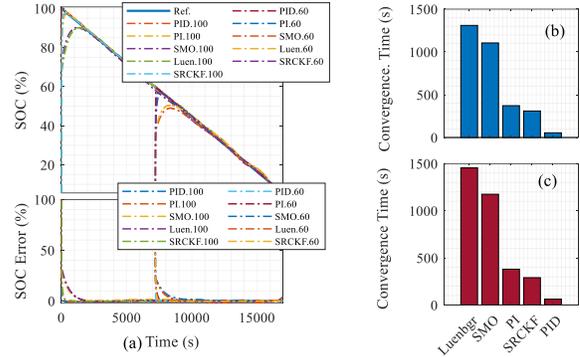

**Fig. 10.** (a) convergence speed results of all methods: comparative convergence time for all methods (b) *CALB L148N58A* cell (c) *INR 21700M50T* cell

### C. Robustness against Measurement Uncertainties

The battery current and voltage sensor-related measurement errors are a common phenomenon in the practical applications of LIBs. That's why it is critical to have a degree of robustness against these uncertainties due to a narrow range for the safe operations of LIBs. In the case of the current, these uncertainties could be the result of a magnetic or electric offset, linearity, or gain sensitivity. According to the principle of the Hall effect, these errors can be modeled with biased noise of nonzero mean. To evaluate this scenario, a biased noise of standard deviation of 0.01A and mean values of 0.5A, 1A, 2.5A, and 5A is injected into the cell's current data. The resulting estimations of each technique are plotted in Fig. 11, and it can be seen that the errors of the SRCKF increase with the value of current uncertainty; however, these current-related measurement uncertainties have minimal effect on the accuracy of observer-based methods.

To investigate the effects of voltage-related uncertainties, a time-damped signal with a mean value of .004V and standard deviation of 0.005V, 0.01V, 0.1V, and 0.2V is injected into the measured cell voltage signal. It can be clearly understood from the results of Fig. 11 that the estimation error increases against high voltage uncertainties, though the value of the error is much lower in advanced observers like the PI and PID. However, the general trend is almost the same for all methods. This is mainly because all ECM-based SOC estimation methods employ the battery terminal voltage as the tracking target. Therefore, it can be said that the filters are sensitive to both current and voltage-related measurement errors, while observers' SOC estimations are generally very robust to current-related errors.

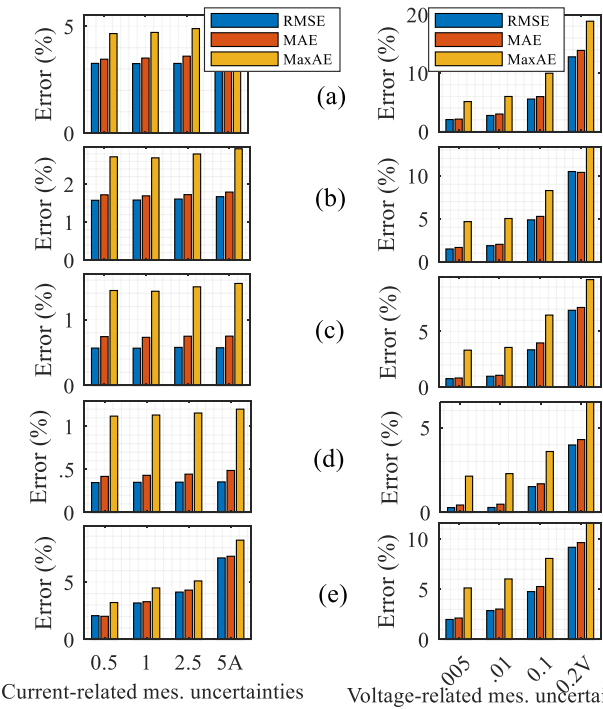

**Fig. 11.** Statistical Error indices against the measurement uncertainties: (a) Luenberger, (b) SM, (c) PI, (d) PID, and (e) SRCKF based SOC estimation

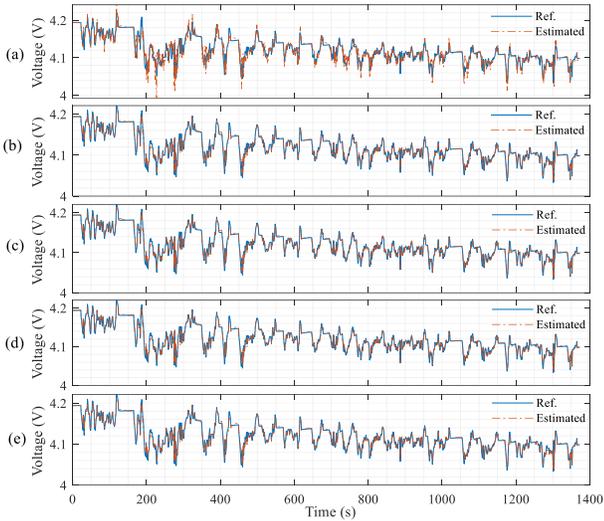

**Fig. 12.** Estimations against sensitivity in (a) $R_{ohm}$ (b) $R_a$ (c) $C_a$ (d) $R_b$ (d) $C_b$

TABLE V
STATISTICAL ERROR INDICES FOR PARAMETERS' SENSITIVITY

| Indictor | $R_{ohm}$ | $R_a$ | $C_a$ | $R_b$ | $C_b$ |
|---|---|---|---|---|---|
| RMSE (mV) | 13.55 | 2.066 | 2.052 | 1.93 | 1.92 |
| MAE (mV) | 10.36 | 1.58 | 1.47 | 1.45 | 1.46 |

*D. Stability*

All the model-based SOC estimation methods are sensitive to the cell parameter values. As the operating environment vary largely in practical situations, it is common to observe changes in cell parameters' value due to the high vulnerability of LIBs to the operating conditions. Such parametric changes can cause a downshift in the performance, accuracy, and stability of the system. To evaluate the effect of these changes on the control observer's performance, a parameter sensitivity analysis is first performed. The analysis is carried out using a FUDS current profile and ±80% of parameters identified value provided as the input-synchronized time-damped sensitivity signals to the SOC estimator. The analysis results are plotted in Fig. 12, and error indices for each parameter's sensitivity are listed in Table V.

It is clear from the results that the SOC estimator's response is almost insensitive to the perturbations of $R_a$, $R_b$, $C_a$, and $C_b$. However, the extent of performance deterioration in the case of ohmic polarization ($R_{ohm}$) is much higher than other battery parameters which easily classifies $R_{ohm}$ as the dominant parameter for the model-based SOC estimation methods. That's why, the effectiveness of each technique is then evaluated against the sensitivity in the dominant parameter $R_{ohm}$, and the estimation results along with error indices are shown in Fig. 13. It can be understood from the results that the Luenberger, SMO and SRCKF show higher deterioration in their estimations, while the PI and PID observes exhibit better tackling against the sensitivity in $R_{ohm}$. However, it can also be understood that the ECM-based methods are somehow considerably sensitive to the changes in the ohmic polarization of LIBs. Therefore, an online estimation of the battery ohmic polarization can enhance the estimation accuracy during practical situations.

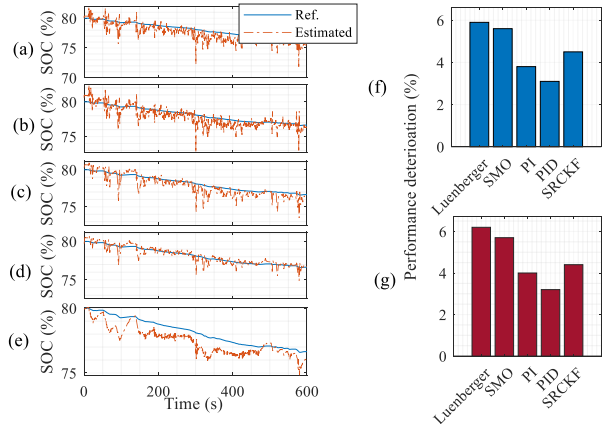

**Fig. 13.** SOC estimations against sensitivity in $R_{ohm}$ (a) Luenberger (b) SMO (c) PI (d) PID (e) SRCKF; Error indices for (f) *L148N58A*, and (g) *INR 21700*

*E. Computational Complexity*

The computational complexity is one of the critical factors affecting the selection criterion when choosing a SOC estimator for commercial utility. In fact, despite obvious disadvantages, coulomb counting techniques are still implemented in many industrial applications due to their low complexity. However, ECM-based SOC methods offer much higher robustness, better accuracy, and advanced control. To evaluate the computational complexity of the established techniques, this study uses the following steps: 1) The battery data for one experiment is recorded in the host PC. 2) Then, the recorded data of a specific period (taken 132000s in this work) is provided directly as input to each method in SIMULINK, and is executed 10 times. 3) Lastly, the SIMULINK profiler function is used to record the computational time for each execution, and the average value is computed for each method. The same procedure is repeated for with added disturbances and non-linearities. A Lenovo desktop with Windows 10 pro, 32GB RAM, and an Intel core i7-6700 CPU is used for this work, and the results are shown in Fig. 14.

The SRCKF shows the highest computational time (>21s) which is one of the major issues of filter-based methods. Among observers, the Luenberger shows a higher consumption time of 8.14s because of having only the P-control, and a high value of P-gain is required to make it properly track the state values of complex LIB systems. The computational time of the SM observer is well reduced to 2.65s due to the additional discontinuous feedback signal. Moreover, the presence of advanced control helps the PI and PID observers in an efficient error convergence within 1.9s and 1.95s, respectively. As the Luenberger observer is not involved with the sign, rate, and history of the error, the added disturbances and nonlinearities do not have much effect on its computational time. However, the computational time for the SM observer increases to a relatively higher value of 7.49s because the added disturbances activate aggressive discontinuous feedback. Similarly, an increase in the value of the computational time of PI and PID observers is also observed; however, the statistics for the PI and PID observers are still much more efficient than the former two.

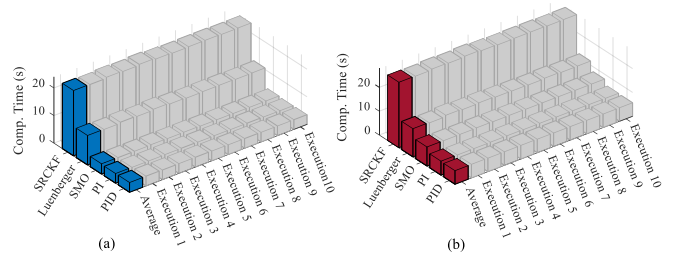

**Fig.14.** The computational time of each SOC estimation techniques against (a) linearized scenario (b) with modeled non-linearities and added disturbances.

## V. Conclusion

This work successfully performed a systematic comparison of control observer-based methods for state estimation of Li-ion batteries (LIBs) considering different practical scenarios. The selected techniques are designed using a $2^{nd}$ order equivalent circuit model (ECM) whose rigorous observability is discussed against multiple scenarios. All established techniques proved a strong candidacy for the state estimation of LIBs; however, the following deductions can made from comparative results. 1) The observer-based methods are capable of estimating accurate SOC, their accuracy is less affected by high temperatures and highly dynamic loadfiles, and they exhibit better compatibility to practical environments as compared to filter-based methods. 2) Among observers, the PID provides the fastest convergence against unexpected shutdown and incorrect initial value. 3) Unlike filter-based methods, all observers show higher robustness against current-related uncertainties while being only sensitive to voltage-related errors. 4) The observers-based method shows sensitivity to the cell ohmic polarization; however, the extent of sensitivity decreases as we move from the Luenberger towards the PI and PID observers. Overall, the observers (especially PI and PID)-based SOC are well suitable for real-time applications due to the improved accuracy, enhanced robustness, faster convergence, and low complexity; however, a framework with a facility for joint estimation of cell ohmic resistance, and active detection of voltage-bias is advised for high-end applications.

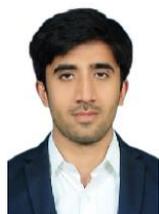

**Muhammad Saeed** (*Graduate Student Member*, IEEE) received his B.E and M.E degrees from the National University of Sciences and Technology, Islamabad, Pakistan, and the Northwestern Polytechnical University, Xi'an, respectively. He is currently working toward the Ph.D. degree in protection and control of Lithium-ion batteries at Chongqing University. From 2018 to 2019, he was with Energy China (CEEC) and Siemens in the power and energy storage sector. His research interests include state estimation, Li-ion batteries, and physics-informed intelligent control.

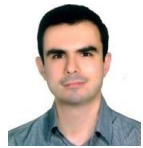

**Arash Khalatbarisoltani** (S'18) received the Ph.D. in electrical engineering from Université du Québec à Trois-Rivières (UQTR), Canada, in 2022. He is a Postdoctoral Researcher with the Department of Mechanical and Vehicle Engineering, Chongqing University, China. His research interests include electrified vehicles and intelligent transport.

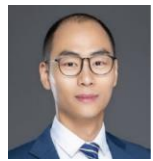

**Zhongwei Deng** (Member, IEEE) received his Ph.D. from Shanghai Jiao Tong University, Shanghai, China, in 2019. He is currently an Associate Professor in the School of Mechanical and Electrical Engineering, University of Electronic Science and Technology of China. His research interests focus on data-driven and mechanism modeling, state estimation, health diagnosis, and second-life of LIBs.

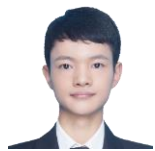

**Wenxue Liu** received the B.E. and M.S. degrees in Automotive Engineering from Chongqing University, Chongqing, China, in 2017 and 2020, respectively. He is currently pursuing the PhD. degree in coupled modeling and state estimation of Li-ion batteries with the College of Mechanical & Vehicle Engineering, Chongqing University.

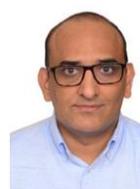

**Faisal Altaf** (Member, IEEE) received the B.Sc. degree in mechatronics engineering from the National University of Sciences and Technology (NUST), Islamabad, Pakistan, in 2004, the M.Sc. degree in electrical engineering from the KTH Royal Institute of Technology, Stockholm, Sweden, in 2011, and the Ph.D. degree in control systems from the Chalmers University of Technology, Gothenburg, Sweden, in 2016. His main focus has been on modeling, estimation, and optimal control design for automotive battery management systems (BMS) since 2011. From 2016 to 2021, he worked as a Lead Control System Engineer at NEVS, Västergötland, Sweden, and a Principal Researcher and a Control System Architect at Volvo Group Trucks Technology, Gothenburg. He is currently working as a Chief Engineer at the Electromobility Department, Volvo Group Trucks Technology, where he is leading research and development on advanced BMS technologies for a range of heavy-duty electric vehicles. His current research interests are at the intersection of traction batteries, control engineering, power electronics, embedded systems, data analytics, and system engineering with a special focus on model- and learning-based controls for automotive applications.

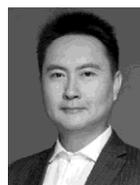

**Shuai Lu** (Member, IEEE) received the B.S.E.E. degree from the Chongqing University, Chongqing, China, in 1997, the M.S.E.E. degree from the University of Wisconsin-Milwaukee, Milwaukee, WI, USA, in 2003, and the Ph.D. degree from the University of Missouri-Rolla, Rolla, MO, USA, in 2007, all in electrical engineering. In February 2007, he joined MTS Systems Corporation, Eden Prairie, MN, USA, where he was the Lead Power Electronics and Motor Drive Engineer for the successful development of the world's first generation of the hybrid electric system for Formula-1 cars in the 2009 race season (also known as KERS: kinetic energy recovery system). In late 2012, he joined Chongqing University as Professor, where he has accomplished numerous industrial research and development projects in various areas of power electronics and motor drives, with the particular focus on the applications in hybrid and electric vehicles, renewable energy power generation systems, and microgrid power conversions.

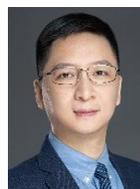

**Xiaosong Hu** Xiaosong Hu (Fellow, IEEE) received the Ph.D. degree in automotive engineering from the Beijing Institute of Technology, Beijing, China, in 2012. From 2010 to 2012, he completed his scientific research and Ph.D. dissertation with the Automotive Research Center, University of Michigan, Ann Arbor, MI, USA. He is currently a Professor with the College of Mechanical and Vehicle Engineering, Chongqing University, Chongqing, China. His research interests include modeling and control of alternative powertrains and energy storage systems.

Dr. Hu was the recipient of numerous prestigious awards/honors, including Web of Science Highly-Cited Researcher by Clarivate Analytics, SAE Environmental Excellence in Transportation Award, SAE/Timken-Howard Simpson Automotive Transmission and Driveline Innovation Award, IEEE ITSS Young Researcher Award, and SAE Ralph Teetor Educational Award. He is an IET Fellow and an AAIA Fellow.